\documentclass[aps,prl,twocolumn,superscriptaddress,footinbib,longbibliography,amsmath,amssymb]{revtex4-2} 

\usepackage{graphicx}
\usepackage{subfigure}
\usepackage{epsfig} 
\usepackage{dcolumn}
\usepackage{bm}
\usepackage{times} 
\usepackage{amsmath}
\usepackage{float}
\usepackage{color}
\usepackage{multirow}
\usepackage[normalem]{ulem}

\begin{document}
\title{Near-field energy transfer between graphene and magneto-optic media}

\author{Gaomin Tang}
\email{gaomin.tang@unibas.ch}
\affiliation{Department of Physics, University of Basel, Klingelbergstrasse 82, CH-4056
Basel, Switzerland}
\author{Lei Zhang}
\affiliation{State Key Laboratory of Quantum Optics and Quantum Optics Devices, Institute
of Laser Spectroscopy, Shanxi University, Taiyuan 030006, China}
\affiliation{Collaborative Innovation Center of Extreme Optics, Shanxi University, Taiyuan
030006, China}
\author{Yong Zhang}
\affiliation{School of Energy Science and Engineering, Harbin Institute of Technology,
Harbin 150001, P. R. China}
\affiliation{Key Laboratory of Aerospace Thermophysics, Ministry of Industry and
Information Technology, Harbin 150001, P. R. China}
\author{Jun Chen}
\email{chenjun@sxu.edu.cn}
\affiliation{State Key Laboratory of Quantum Optics and Quantum Optics Devices, Institute
of Theoretical Physics, Shanxi University, Taiyuan 030006, China}
\author{C. T. Chan}
\affiliation{Department of Physics and Institute for Advanced Study, The Hong Kong
University of Science and Technology, Hong Kong, China}

\bigskip

\begin{abstract}
  We consider the near-field radiative energy transfer between two separated parallel
  plates: graphene supported by a substrate and a magneto-optic medium.
  We first study the scenario in which the two plates have the same temperature. 
  An electric current through the graphene gives rise to nonequilibrium fluctuations and
  induces energy transfer. 
  Both the magnitude and direction of the energy flux can be controlled by the electric
  current and an in-plane magnetic field in the magneto-optic medium. 
  This is due to the interplay between the nonreciprocal photon occupation number
  in the graphene and nonreciprocal surface modes in the magneto-optic plate.
  Furthermore, we report that a tunable thermoelectric current can be generated in the
  graphene in the presence of a temperature difference between the two plates. 
\end{abstract}

\maketitle

{\it Introduction.--}
Nonreciprocity is attracting substantial interest in plasmonics and radiative energy
harvesting. In magneto-optic materials, a magnetic field breaks time-reversal symmetry,
which results in nonreciprocal electromagnetic surface waves. 
Owing to the broken reciprocity, various novel near-field heat transfer phenomena have
been reported, including photon thermal Hall effect~\cite{nonrecip_Hall}, persistent heat
current~\cite{nonrecip_persistent,nonrecip_18}, thermal
magnetoresistance~\cite{control_magnetic_15,control_magnetic_17,
control_magnetic_18,WSM_radiate2}, thermal rectification~\cite{nonrecip_rect}, and Casimir
heat engine~\cite{propulsion_21, nonrecip_lateralCasimir}. 

In graphene, nonreciprocity can be induced by applying an electric current~\cite{drift_15,
drift_16, drift_18, drift_20, Papaj_20}. Interesting properties of graphene plasmons such
as negative Landau damping~\cite{drift_17} and Fizeau drag~\cite{Fizeau1, Fizeau2} have
been studied. 
Current-biased graphene is in a nonequilibrium state, which leads to a finite photonic
chemical potential~\cite{photon_noise_RMP} that depends on the in-plane wave vector
${\bf q}$ for the thermal electromagnetic radiation.
The occupation number of radiative photons at angular frequency $\omega$ and
electronic temperature $T$ becomes nonreciprocal, 
$n(\omega) = [e^{\hbar(\omega - {\bf q} \cdot {\bf v}_d)/k_B T} -1]^{-1}$, where 
${\bf v}_d$ is the drift velocity of the electric current~\cite{VP08_PRB, VP11_PRL,
VP11_PRB, drift1}. 
Regarding regulation of the photonic chemical potential, a p-n junction with a voltage
bias has been reported to enable solid-state cooling in the near-field
regime~\cite{photon_potential_15, photon_potential_16, photon_potential_19}. 

\begin{figure}
\centering
\includegraphics[width=\columnwidth]{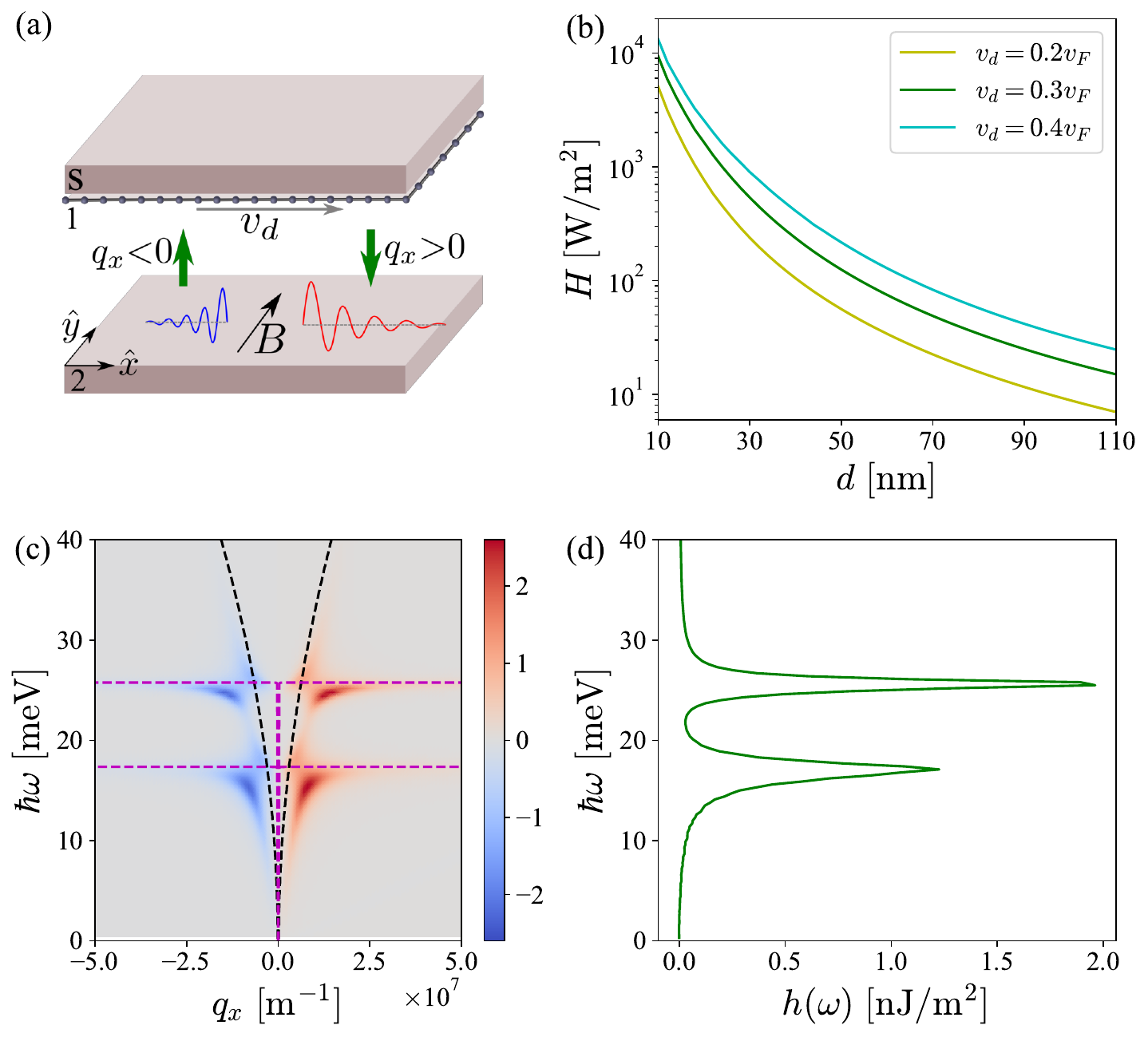} \\
\caption{(a) Schematic plot of the near-field energy transfer between graphene (1)
  supported by a substrate (s) and a magneto-optic medium plate (2) with air gap $d$. 
  In the presence of an electric current with drift velocity $v_d$ through the graphene, a
  net energy flux is transferred even when there is no temperature difference between the
  two plates. 
  Its magnitude and direction can be modulated by the electric current and the magnetic
  field $B$ applied to the magneto-optic medium. 
  (b) Energy flux $H$ versus $d$ at different $v_d$ with $B=0$ and $T=300\,$K.
  (c) Energy transmission function ${\cal Z}$ (in units of meV) at $v_d=0.3v_F$,
  $d=20\,$nm, and $q_y=0$ against $q_x$ and $\hbar\omega$. 
  The black and magenta dashed lines are the dispersions of graphene plasmons and the
  surface modes of InSb at $B=0$ in the absence of damping, respectively. 
  (d) Spectrum $h(\omega)$ at $v_d=0.3v_F$ and $d=20\,$nm. }
\label{fig1}
\end{figure}

In this Letter, we study the near-field energy transfer between two parallel plates:
graphene supported by a substrate and a magneto-optic plate [see Fig.~\ref{fig1}(a)]. In
the presence of an electric current through the graphene, a net energy is transferred
between the plates when they have an equal electronic temperature. 
An in-plane magnetic field perpendicular to the electric current is applied to
the magneto-optic medium. 
In the absence of the magnetic field, the energy flux flows from the graphene to the
magneto-optic medium in the absence of a temperature difference between the plates.
Remarkably, the energy flux direction can be controlled by the magnetic field due to the
nonreciprocal surface modes of the magneto-optic medium.
Furthermore, a tunable thermoelectric current can be generated in the graphene with a
temperature difference between the two plates.  

{\it Formalism.--}
The system is schematically shown in Fig.~\ref{fig1}(a) where the substrate, graphene, and
magneto-optic plate are denoted by indices $s$, $1$, and $2$, respectively. 
Within the framework of fluctuational electrodynamics~\cite{Rytov,PvH}, the near-field
radiative energy flux $H$ on the magneto-optic plate is given by~\cite{many12PRL,
many14PRA, many17PRB, review20}
\begin{equation} \label{H}
  H =\int_0^{\infty} \frac{d\omega}{2\pi} \int_{c|{\bf q}|{>}\omega} 
  \frac{d^2{\bf q}}{4\pi^2} \hbar\omega[ n_{s2}\xi_{s2}(\omega,{\bf q}) +
  n_{12}\xi_{12}(\omega,{\bf q}) ],
\end{equation}
where ${\bf q}=(q_x, q_y)$ is the in-plane wave vector and $\omega$ is the angular
frequency. Here, $n_{ij}\equiv n_i-n_j$ is the occupation difference of the photons
radiated from object $i$ and $j$, with $i,j\in \{s,1,2\}$. 
Photonic transmission coefficients $\xi_{s2}$  and $\xi_{12}$ with air gap separation
$d$ are expressed as
\begin{align}
  \xi_{s2} &= 4|\tau_1|^2{\rm Im}(\rho_s){\rm Im}(\rho_2)e^{-2k_z d} |u_{s1,2}|^2
  |u_{s,1}|^2 , \\
  \xi_{12} &= 4{\rm Im}(\rho_{s1}){\rm Im}(\rho_2)e^{-2k_z d} |u_{s1,2}|^2 -\xi_{s2},
  \label{xi12}
\end{align}
with $k_z = \sqrt{|{\bf q}|^2 - (\omega/c)^2}$, 
$u_{s1,2}=(1-\rho_{s1} \rho_2 e^{-2 k_z d})^{-1}$, and $u_{s,1}=(1-\rho_s \rho_1)^{-1}$. 
In the above, $\rho_i$ is the reflection coefficient for the $p$-polarized mode of object
$i$, and $\tau_1$ is the transmission coefficient of the graphene which is treated as a
thin film with finite thickness~\cite{vakil2011, lim2013}. 
In addition, $\rho_{s1}=\rho_1 + \tau_1^2 \rho_s u_{s,1}$ is the reflection coefficient of
the graphene supported by the substrate. 
Details of these coefficients are provided in the Supplemental Material~\cite{SM}. 

We now consider the scenario in which an electric current is applied through the graphene
and all the objects have the same temperature $T$. 
In this case, $n_{s2}$ vanishes, and Eq.~\eqref{H} becomes
\begin{equation} \label{H1}
  H =\int_0^{\infty} \frac{d\omega}{2\pi} \int_{c|{\bf q}|{>}\omega} 
  \frac{d^2{\bf q}}{4\pi^2} \hbar\omega n_{12}(\omega, q_x) \xi_{12}(\omega,{\bf q}) .
\end{equation}
Without loss of generality, the electric current-induced drift velocity $v_d$ in the
graphene is along the positive direction of the $x$ axis. 
The photon occupation number difference between the graphene and magneto-optic medium is
thus~\cite{review07, VP08_PRB, VP11_PRL, VP11_PRB, Cherenkov13, drift1}
\begin{equation} \label{n12}
  n_{12}(\omega, q_x) = [e^{\hbar(\omega - q_x v_d)/k_B T} -1]^{-1}
  - [e^{\hbar\omega/k_B T} -1]^{-1} .
\end{equation}
Equation~\eqref{n12} implies that the electric current results in a finite energy flux
even in the absence of a temperature difference between the two plates. 
From Eqs.~\eqref{H1} and \eqref{n12}, a positive energy flux indicates that it flows from
the graphene to the magneto-optic medium and inversely for a negative flux. 
In this work, we consider a gap separation $d$ of not less than $10\,$nm so that the
contribution from $\omega {<} q_x v_d$, which is only sizable with a subnanometer
separation~\cite{VP11_PRL, VP11_PRB}, can be neglected.
For positive $q_x$, the energy transfer is from the graphene to the magneto-optic medium
since $n_{12}$ is positive under $\omega {>} q_x v_d$. 
For negative $q_x$, $n_{12}$ is negative,  
and the energy transfer direction is opposite to that for positive $q_x$. 
We define the energy transmission function of the energy flux in Eq.~\eqref{H1} as
\begin{equation}
  {\cal Z}(\omega, {\bf q}) = \hbar\omega n_{12}(\omega, q_x) \xi_{12}(\omega,{\bf q}),
\end{equation}
which gives the energy transfer at given $\omega$ and ${\bf q}$. The energy flux spectrum
$h(\omega)$ is defined through $H = \int_0^{\infty} d\omega h(\omega)/2\pi$. 

For a three-dimensional object in the presence of magnetic field $B$, its dielectric
tensor has off-diagonal elements due to the cyclotron motion at frequency
$\omega_c=eB/m^*$, where $m^*$ is the effective electron mass. 
Usually, the off-diagonal elements are negligible.
In magneto-optic materials which have low carrier density and small effective mass, the
cyclotron motion frequency $\omega_c$ can be comparable to the plasma frequency, which
leads to large off-diagonal elements. 
Because of this, the surface modes of magneto-optic materials are nonreciprocal in the
Voigt configuration. To study the interplay of the nonreciprocal effects from the
photon occupation number and the surface modes, the magnetic field is applied along
the $y$ direction so that the dielectric tensor reads
\begin{equation} \label{tensor}
  \bar{\bar{\epsilon}}(\omega) = 
  \begin{bmatrix}
    \epsilon_d & 0 & i\epsilon_a \\
    0 & \epsilon_p & 0 \\
    -i\epsilon_a & 0 & \epsilon_d
  \end{bmatrix} ,
\end{equation}
with $\epsilon_a=
\epsilon_{\infty}\omega_c\omega_p^2/\{\omega[(\omega+i\gamma)^2-\omega_c^2]\}$ using the
Drude model. 
Here, $\epsilon_\infty$ is the high-frequency dielectric constant, $\omega_p$ is the
plasma frequency, and $\gamma$ describes the free-carrier damping constant. 
The expressions of $\epsilon_d$ and $\epsilon_p$ are given in the Supplemental
Material~\cite{SM}. 
The dispersion relation of $p$-polarized surface modes with wave vector $q_x$
is~\cite{BinHu15}
\begin{equation} \label{dispersion}
  \epsilon_v \sqrt{(\omega/c)^2 -q_x^2} +\sqrt{\epsilon_v (\omega/c)^2 -q_x^2}
  -i\epsilon_a q_x /\epsilon_d =0 ,
\end{equation}
with 
$\epsilon_v = \epsilon_d-\epsilon_a^2/\epsilon_d$. This expression explicitly indicates
that $\epsilon_a$ gives rise to the nonreciprocal dispersion. 

In magnetic Weyl semimetals, Weyl-node separation in momentum space leads to the anomalous
Hall effect~\cite{WSM_AHE_18np, WSM_AHE_19PRB, WSM_AHE_20nc} such that there are
off-diagonal components in the dielectric tensor. Compared to magneto-optic
materials, magnetic Weyl semimetals intrinsically support nonreciprocal surface
polaritons~\cite{WSM_SPP16, Kotov18}.  Therefore, the magneto-optic medium can be
replaced by a magnetic Weyl semimetal such that an external magnetic field is not
needed~\cite{WSM_radiate1, WSM_radiate2, WSM_radiate3, WSM_radiate4, GT_WSM}. 

\begin{figure}
\centering
\includegraphics[width=\columnwidth]{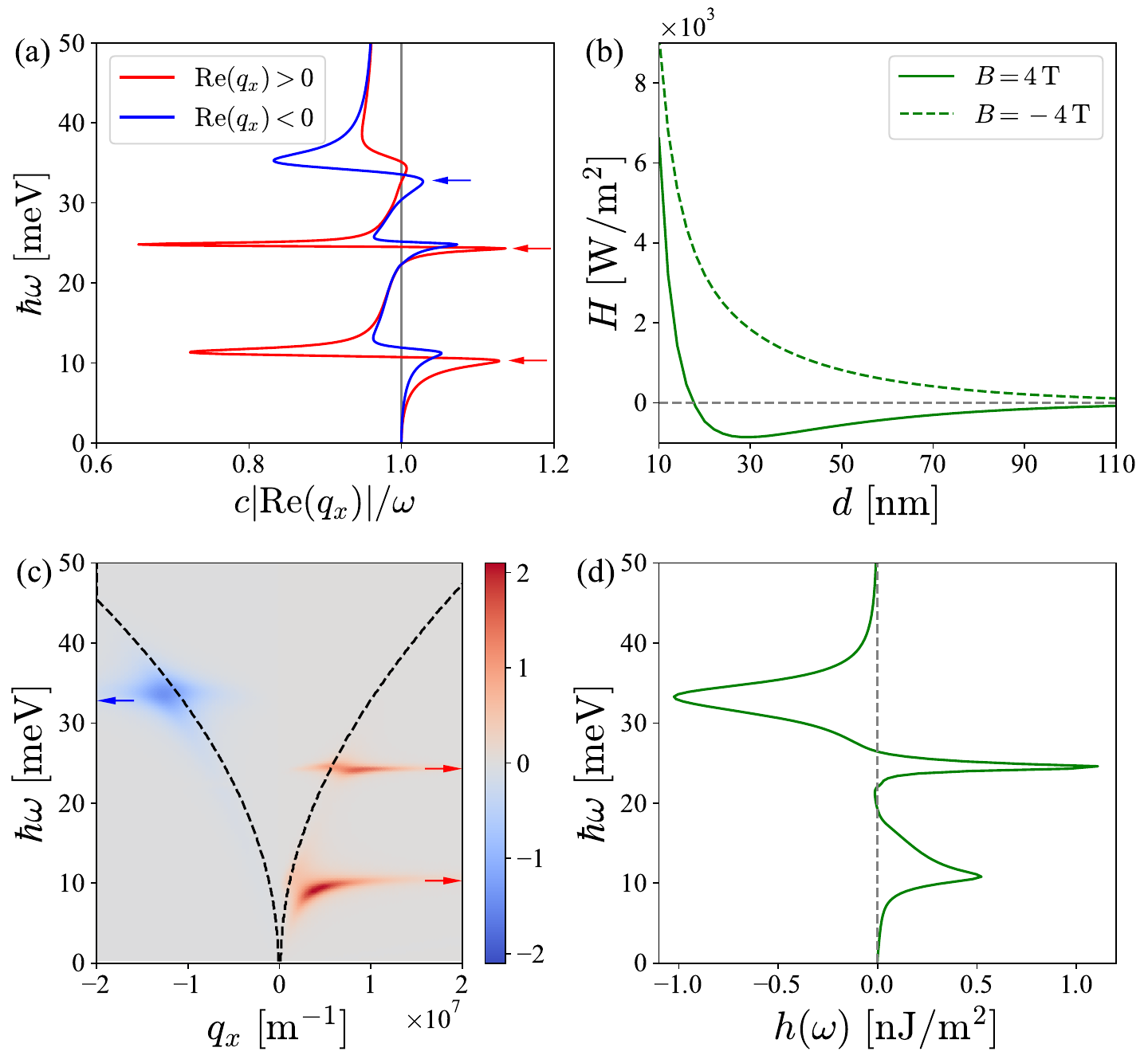} \\
\caption{(a) Dispersion of the surface modes of InSb in the Voigt configuration at
  $B=4\,$T. 
  (b) Energy flux $H$ versus separation $d$ at $B=4\,$T and $B=-4\,$T with $v_d=0.3v_F$
  and $T=300\,$K. 
  (c) Energy transmission function ${\cal Z}$ (in units of meV) at $B=4\,$T, $d=50\,$nm,
  and $q_y=0$ against $q_x$ and $\hbar\omega$. The black dashed lines are dispersions of
  the graphene plasmons. The energies $\hbar\omega$ indicated by the arrows are the same
  as the corresponding energies indicated in (a). 
  (d) Spectrum $h(\omega)$ at $B=4\,$T and $d=50\,$nm. }
\label{fig2}
\end{figure}

In the numerical calculation, we choose hexagonal boron nitride as the substrate.
The chemical potential of the graphene is $0.1\,$eV. The drift velocity in the graphene is
set to around $0.3v_F$ so that the nonreciprocity of the graphene plasmons can be
neglected~\cite{drift_15}. This can be seen from the black dashed lines in
Fig.~\ref{fig1}(c). The magneto-optic medium is chosen to be InSb. 
The contribution from the term $\xi_{s2}$ in $\xi_{12}$ given by Eq.~\eqref{xi12} is
negligible due to the mismatch between the surface modes from the hexagonal boron nitride
and InSb.
Below, we separately discuss the scenarios in the absence and presence of a magnetic field
at $T=300\,$K.

{\it Energy transfer at $B {=} 0$.--}
Figure~\ref{fig1}(b) shows the energy flux versus gap separation $d$ in the absence of
a magnetic field. The positive value of the energy flux is explained as follows.
Under $B=0$, the surface polaritons of the InSb plate are reciprocal, as is the
transmission coefficient $\xi_{12}$.
Because $n_{12}(\omega, q_x) {>} -n_{12}(\omega, -q_x)$ for $\omega {>} q_x v_d$ with 
$q_x {>} 0$, the energy transfer of the positive $q_x$ dominates over that of
the negative $q_x$, and the net energy flux is from the graphene to the InSb plate. 
This argument is supported by the energy transmission function ${\cal Z}$ in
Fig.~\ref{fig1}(c) and the spectrum $h(\omega)$ in Fig.~\ref{fig1}(d).
The two positive peaks in the spectrum correspond to the two
surface-polariton frequencies of InSb and are due to the difference between the
contributions from positive and negative $q_x$. 
Figure~\ref{fig1}(b) shows that the energy flux decreases with
increasing separation $d$, which is due to the evanescent nature of the surface
modes.


{\it Energy transfer at finite $B$.--}
We now consider the situation in which a magnetic field ${\bf B}{=}B\hat{y}$ is applied to
the magneto-optic plate. Positive and negative magnetic fields $B$ are defined as being
along the positive and negative directions of the $y$ axis, respectively.
For $B{>}0$, 
the dispersion of the surface polaritons in the magneto-optic plate is redshifted for
positive $q_x$ and experiences a blueshift for negative $q_x$ compared to the dispersion
in the absence of a magnetic field~\cite{nonrecip_torque19}.
This scenario is opposite for $B{<}0$. 
The surface polaritons of the InSb plate for positive and negative ${\rm Re}(q_x)$ at
$B=4\,$T are shown in Fig.~\ref{fig2}(a). Here, $q_x$ is complex because the damping
constants are taken into account in Eq.~\eqref{dispersion}. The polaritons at ${\rm
Re}(q_x){<}0$ are more strongly damped than those at ${\rm Re}(q_x){>}0$. 
For the case with reciprocal surface modes and photonic transmission coefficients, the
energy transfer of $n_{12}(\omega, q_x)$ with $q_x v_d{>}0$ dominates over that of
$n_{12}(\omega, -q_x)$, which leads to the net energy flux flowing out of the graphene
sheet. 
However, because of the nonreciprocal surface polaritons of the magneto-optic medium, the
photonic transmission coefficients are different for wave vectors of $(q_x, q_y)$ and
$(-q_x, q_y)$. This enables the net energy flux direction to be changed by a magnetic
field. 

Figure~\ref{fig2}(b) shows the energy flux versus the gap separation at $B= 4\,$T
(solid line) and $B=-4\,$T (dashed line) with $v_d=0.3v_F$. At $B=4\,$T, the energy
flux is negative when the gap separation is larger than about $18\,$nm. 
To understand this, we show the energy transmission function in Fig.~\ref{fig2}(c) and the
energy flux spectrum $h(\omega)$ in Fig.~\ref{fig2}(d) at $B=4\,$T and $d=50\,$nm.
The energies $\hbar\omega$ around which the energy transfer is prominent are indicated by
arrows in Fig.~\ref{fig2}(c). 
The arrows correspond to the peaks in Fig.~\ref{fig2}(d). 
The contributions from the positive and negative $q_x$ give rise to the positive and
negative peaks, respectively. 
The negative peak is broader than the two positive peaks, which originates from the
nonreciprocity of the surface polaritons of InSb. Therefore, the increased polariton
damping at negative $q_x$ and decreased damping at positive $q_x$ give rise to the
negative energy flux at $B=4\,$T and $d\gtrsim 18\,$nm. The energy flux is positive at
$B=4\,$T and $d\lesssim 18\,$nm since the nonreciprocity of the photon occupation from the
graphene increases with increasing magnitude of wave vector $q_x$. 
Thus, the interplay between the nonreciprocal photon occupation number and the
nonreciprocal surface modes governs the net energy transfer direction. 
For $B=-4\,$T, surface polariton broadening of InSb occurs at positive $q_x$ such that the
synergy from the nonreciprocal effects of the photon occupation number and the surface
polaritons gives rise to the positive energy flux in Fig.~\ref{fig2}(b).

\begin{figure}
\centering
\includegraphics[width=\columnwidth]{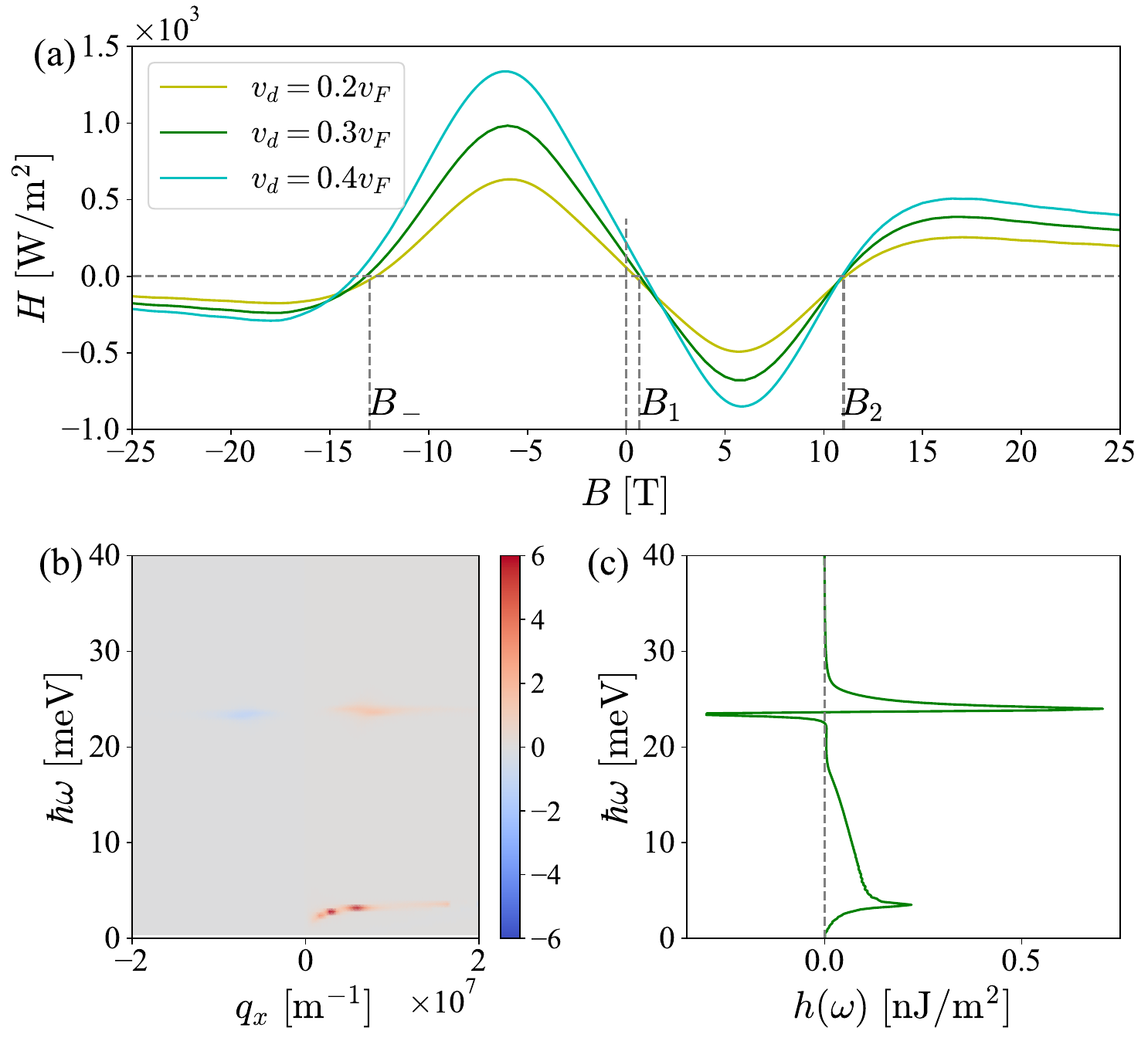} \\
\caption{(a) Energy flux $H$ versus magnetic field $B$ at different drift velocities $v_d$
  with $d=50\,$nm and $T=300\,$K. 
  (b) Energy transmission function ${\cal Z}$ at $q_y=0$, $B=18\,$T, $v_d=0.3v_F$, and
  $d=50\,$nm against $q_x$ and $\hbar\omega$.
  (c) Spectrum $h(\omega)$ at $B=18\,$T and $v_d=0.3v_F$. } 
\label{fig3}
\end{figure}

The energy flux versus magnetic field $B$ behavior at $d=50\,$nm is displayed in
Fig.~\ref{fig3}. 
The energy flux vanishes at certain magnetic fields, which are denoted as $B_-$, $B_1$ and
$B_2$ for $v_d=0.3v_F$ . 
Under a small magnetic field ($0{<}B{<}B_1$), the nonreciprocity of the surface modes of
InSb is weak such that the energy transfer direction is dominated by the nonreciprocity of
the photon occupation in the graphene and is from the graphene to the InSb plate. 
With increasing magnetic field, the energy flux first becomes negative ($B_1{<}B{<}B_2$)
and then positive ($B{>}B_2$). The negative value of the energy flux can be explained in
the same way as that at $B=4\,$T in Fig.~\ref{fig2}(b). 
For $B{\leq}0$, with increasing magnetic field magnitude, the energy flux first
increases and then decreases, finally becoming negative. 
The region of positive energy flux can be explained similarly to that at $B=-4\,$T in
Fig.~\ref{fig2}(b). 

When the magnetic field is sufficiently large, the surface polaritons in InSb become
highly nonreciprocal. For a large positive magnetic field, the surface waves at $q_x{<}0$
are strongly damped such that their contribution to the energy transfer is suppressed.
Therefore, the energy transfer contribution from $q_x{>}0$ dominates over that from
$q_x{<}0$ above a certain magnetic field [$B_2$ at $v_d=0.3v_F$ in Fig.~\ref{fig3}(a)],
which leads to a net energy flow from the graphene to the InSb plate. This explanation is
supported by Figs.~\ref{fig3}(b) and \ref{fig3}(c). For a large magnetic field along the
negative $y$ axis [$B{<}B_-$ in Fig.~\ref{fig3}(a)], the contribution from $q_x{<}0$ is
dominant, which leads to a negative energy flux. 

In the above, we have discussed the scenario in which the in-plane magnetic field is
perpendicular to the electric current. When the magnetic field is parallel to the electric
current, the surface polaritons of the magneto-optic medium are reciprocal along the
direction of the electric current. In this case, the net energy transfer is from the
graphene to the magneto-optic medium since there is no interplay between the nonreciprocal
effects. 

Notably, our findings do not violate the second law of thermodynamics. Energy could
possibly be transferred from one body to another at a higher temperature if there is an
external agent. This can be applied to our case in which the two bodies have the same
temperature, as an external agent is needed to maintain the electric current in the
graphene.  

\begin{figure}
\centering
\includegraphics[width=\columnwidth]{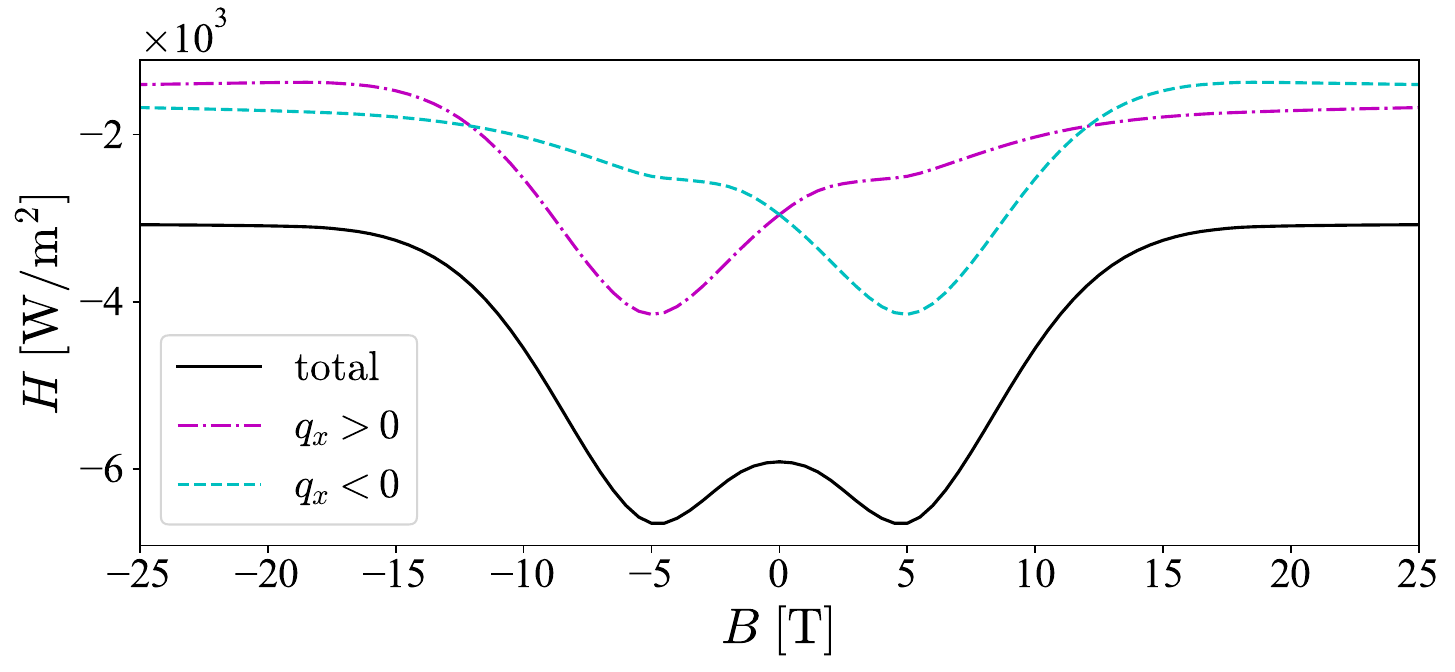} \\
\caption{Heat flux $H$ versus magnetic field $B$ at $T_1=300\,$K, $T_2=330\,$K, and
  $d=50\,$nm (solid line). The contributions to the heat flux from positive $q_x$
  (dash-dotted line) and negative $q_x$ (dashed line) are plotted separately. The negative
  heat flux indicates that the heat flows from the InSb plate to the graphene. }
\label{fig4}
\end{figure}

{\it Thermoelectric effect.--}
The phenomena discussed above imply a thermoelectric effect. We consider the same setup
with a temperature difference between the two plates, and no external electric current is
applied to the graphene. 
In this case, we have $n_{s2}=n_{12}$, and the heat flux is expressed by
\begin{equation}
  H =\int_0^{\infty} \frac{d\omega}{2\pi} \int_{c|{\bf q}|{>}\omega} 
  \frac{d^2{\bf q}}{4\pi^2} \hbar\omega \, n_{12}(\omega) 
  [\xi_{12}(\omega,{\bf q})+\xi_{s2}(\omega,{\bf q})] ,
\end{equation}
with $n_{12}(\omega) = \big(e^{\hbar\omega/k_B T_1} -1\big)^{-1} - \big(e^{\hbar\omega/k_B
T_2} -1\big)^{-1}$. Here, $T_1$ and $T_2$ are the temperatures of the graphene plate and
InSb, respectively. 
When a magnetic field is applied along the $y$ axis in the InSb plate, the heat transfer
rates at momenta $q_x$ and $-q_x$ are different such that the electrons in the graphene of
momenta $q_x$ and $-q_x$ have different occupations [see Fig.~\ref{fig4}]. 
When the impurity scattering in the graphene is weak, this occupation difference can
induce an electric current along the $x$ direction in the graphene due to its high
electron mobility. The magnitude and direction of the induced electric current can be
controlled by the magnetic field. Thus, the near-field thermoelectric effect can be
achieved. 
In the absence of a magnetic field, the in-plane isotropy is preserved such that an
in-plane electric current cannot be generated.
The thermoelectric effect is analogous to the Casimir heat engine in
Refs.~\cite{propulsion_21, nonrecip_lateralCasimir}, where the thermal energy is converted
into mechanical work. 
Notably, the thermoelectric effect in this work is different from that in the traditional
near-field thermophotovoltaic cell using a p-n junction~\cite{TPV-review18}.


To conclude, we have studied the near-field energy transfer between graphene and
magneto-optic media. The energy transfer is induced by the electric current through the
graphene in the absence of a temperature difference. Its direction and magnitude can be
tuned by the electric current and the in-plane magnetic field in the magneto-optic
medium. The tunability of the direction is due to the interplay between the nonreciprocal
photon occupation number from the graphene and the nonreciprocal surface modes of
the magneto-optic medium. 
We have also proposed a form of the near-field thermoelectric effect in which the electric
current can be generated using nonreciprocal surface modes. Our work paves a new route
toward nanoscale energy/thermal management and harvesting using nonreciprocity. 

\bigskip

\begin{acknowledgments}
G.T. acknowledges financial support from the Swiss National Science Foundation (SNSF) and
the NCCR Quantum Science and Technology.
L.Z. and J.C. acknowledge support from the National Natural Science Foundation of China
(Grants No. 12174231, 12074230, 12047571), National Key R\&D Program of China under Grants
No. 2017YFA0304203, 1331KSC, and Shanxi Province 100-Plan Talent Program.
Y.Z. acknowledges support from the National Natural Science Foundation of China (Grant No.
52076056). 
C.T.C. acknowledges support from the Hong Kong RGC (16303119). 
\end{acknowledgments}

\bibliography{bib_heat_radiation}{}

\clearpage

\begin{center}
  \large{\bf{Supplemental Material for ``Near-field energy transfer between graphene and
  magneto-optic media"}}
\end{center}

\section*{Graphene with substrate}
In the presence of an electric current along the $x$ axis, the polarization function 
of a bare graphene sheet at angular frequency $\omega$ and in-plane wave vector 
${\bf q}=(q_x,q_y)$ is expressed as~\cite{drift_17_c1, drift1}
\begin{align}
  \Pi(\omega,q_x,q_y) &= \frac{\mu(T)}{(\pi\hbar v_F)^2} \int_0^{2\pi} d\theta
  \frac{1}{(1-\cos\theta v_d/v_F)^2} \times \notag \\
  & \frac{q_x(\cos\theta-v_d/v_F) + q_y\sin\theta}{(\hbar\omega+i\gamma_g)/(\hbar
  v_F)-q_x\cos\theta-q_y\sin\theta} ,
\end{align}
with $\mu(T)=2k_BT\ln [2\cosh(\frac{\mu_g}{2k_BT})]$.
Here, $T$ is the temperature, $\mu_g$ the chemical potential, $\gamma_g$ the damping
parameter, and $v_d$ the drift velocity. The Fermi velocity is $v_F=10^6\,$m/s. The sheet
conductivity is given by 
\begin{equation}
  \sigma_g(\omega,q_x,q_y)= \frac{ie^2\omega}{q^2} \Pi(\omega,q_x,q_y) ,
\end{equation}
with $q^2=q_x^2+q_y^2$.
In the absence of the electric current ($v_d=0$) and in the long-wavelength limit
($q\rightarrow 0$), we can recover the Drude conductivity with intraband
contribution~\cite{JS21}, 
\begin{equation}
  \sigma_{\rm intra}(\omega) = \frac{i\omega}{(\omega+i\gamma_g/\hbar)^2} \frac{2e^2
  k_BT}{\pi\hbar^2}\ln\left[ 2\cosh\Big( \frac{\mu_g}{2k_BT} \Big) \right] .
\end{equation}

\begin{figure}
\centering
\includegraphics[width=2.3in]{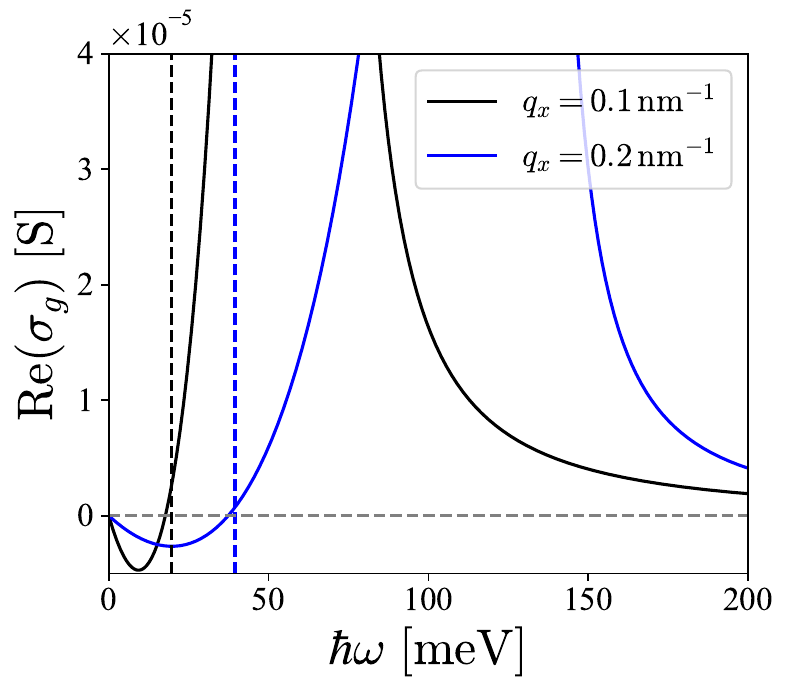} \\
\caption{Real part of the graphene surface conductivity ${\rm Re}(\sigma_g)$ versus
  $\hbar\omega$ for different $q_x$ and $q_y=0$ with electric current along the $x$ axis. 
  Here, $\mu_g=0.1\,$eV, $\gamma_g=3.7\,$meV, $v_d=0.3v_F$, and $T=300\,$K. 
  The vertical dashed lines denote $\omega=q_x v_d$ for the corresponding $q_x$. }
\label{figS1}
\end{figure}

Figure~\ref{figS1} shows the real part of the graphene conductivity ${\rm Re}(\sigma_g)$
for different ${\bf q}=(q_x,0)$ with the electric current along the $x$ axis. One can see
that ${\rm Re}(\sigma_g)$ becomes negative for frequencies below around $q_x v_d$. 
This negative conductivity region leads to the optical gain and negative Landau damping,
which were discussed in bilayer graphene~\cite{drift_17, drift_20}. 
Here, we choose the model in Ref.~\cite{drift_17_c1}, which was generalized to the case of
an arbitrary in-plane wave vector~\cite{drift1}, to describe the polarization function.
In Ref.~\cite{drift_17_c2}, it has been shown that the negative-conductivity region can be
obtained using different models given in Refs.~\cite{drift_15, drift_16, drift_17,
drift_17_c1}. 

We treat the bare monolayer graphene as a thin film with finite thickness
($\Delta=0.3\,$nm), and its dielectric function thus reads~\cite{vakil2011, lim2013}
\begin{equation}
  \epsilon_1 = 1+ \frac{i\sigma_g}{\epsilon_0\omega\Delta} .
\end{equation}
In this work, we only consider $p$ polarized mode which dominates the radiative energy
flux. The ordinary vacuum-graphene Fresnel reflection coefficient is 
\begin{equation}
  r_1 = \frac{\epsilon_1\beta_0-\beta_1}{\epsilon_1\beta_0+\beta_1} ,
\end{equation}
where $\beta_0 = \sqrt{k_0^2 -q^2}$ and $\beta_1 = \sqrt{\epsilon_1 k_0^2 -q^2}$ with
$k_0=\omega/c$. 
The transmission coefficients at the interfaces of vacuum-graphene ($t_1$) and
graphene-vacuum ($\bar{t}_1$) are given by
\begin{equation}
  t_1 = \frac{2\sqrt{\epsilon_1}\beta_0}{\epsilon_1\beta_0+\beta_1}, \qquad 
  \bar{t}_1 = \frac{2\sqrt{\epsilon_1}\beta_1}{\epsilon_1\beta_0+\beta_1} ,
\end{equation}
respectively. 
The reflection and transmission coefficients of the graphene with finite thickness
$\Delta$ are, respectively, given by~\cite{many12PRL, many14PRA}
\begin{equation}
  \rho_1 = r_1 \frac{1- e^{2i\beta_1 \Delta}}{1- r_1^2 e^{2i\beta_1 \Delta}}, \quad
  \tau_1 = \frac{t_1 \bar{t}_1 e^{i\beta_1 \Delta}}{1- r_1^2 e^{2i\beta_1 \Delta}}. 
\end{equation}

We now consider hexagonal boron nitride (hBN) as the substrate. The dielectric tensor of
hBN with the optical axis along the $x$ axis is~\cite{GhBN_15}
\begin{equation}
  \bar{\bar{\epsilon}}_{\rm hBN}(\omega) = 
  \begin{bmatrix}
    \epsilon_{\perp} & 0 & 0 \\
    0 & \epsilon_{\perp} & 0 \\
    0 & 0 & \epsilon_{\parallel} 
  \end{bmatrix} ,
\end{equation}
where
\begin{equation}
  \epsilon_m = \epsilon_{\infty,m} \left( 1+ \frac{\omega_{{\rm LO},m}^2-\omega_{{\rm
  TO},m}^2}{\omega_{{\rm TO},m}^2 -\omega^2 - i\gamma_m\omega} \right)
\end{equation}
with $m=\perp, \parallel$. The parameters are given in Table~\ref{tab:1}.
We assume that the hBN substrate is infinitely thick. The reflection coefficient at the
vacuum-hBN interface is 
\begin{equation}
  \rho_s = \frac{\beta_0\epsilon_\perp -\beta_s}{\beta_0\epsilon_\perp +\beta_s} ,
\end{equation}
with $\beta_s=\sqrt{\epsilon_\perp k_0^2 -\epsilon_\perp q^2 /\epsilon_\parallel}$. 

The reflection coefficient at the interface between vacuum and the graphene-covered hBN
is~\cite{many14PRA} 
\begin{equation} \label{rho_s1}
  \rho_{s1}=\rho_1 + \tau_1^2 \rho_s u_{s,1},
\end{equation}
with $u_{s,1} = (1-\rho_s \rho_1 e^{2i\beta_0 \delta})^{-1}$.
Here, $\delta$ is the effective distance between the graphene and the substrate and is set
to zero in this work. 

Treating the graphene layer with vanishing thickness and using the transfer-matrix method,
the reflection coefficient of $p$ polarized mode for graphene-covered hBN is alternatively
given by~\cite{graphene_plasmon} 
\begin{equation} \label{rho_s1_prime}
  \rho'_{s1} = \frac{\beta_0\epsilon_\perp -\beta_1 +
  \beta_0\beta_1\sigma_g/(\omega\epsilon_0)}{\beta_0\epsilon_\perp +\beta_1 +
  \beta_0\beta_1\sigma_g/(\omega\epsilon_0)} .
\end{equation}
From Eq.~\eqref{rho_s1}, the dispersion of the surface plasmon polariton of the graphene
supported by the substrate hBN is expressed as
\begin{equation}
  \beta_0\epsilon_\perp +\beta_1 + \beta_0\beta_1\sigma_g/\omega = 0. 
\end{equation}
For near-field heat radiation, it was shown that Eqs.~\eqref{rho_s1} and
\eqref{rho_s1_prime} provide identical results~\cite{lim2013}. 

\begin{table}[t] 
  \begin{center} 
    \begin{tabular}{|c||c|c|c|c|}
      \hline 
        & $\epsilon_{\infty,m}$ & $\omega_{{\rm TO},m}$ (${\rm rad}/{\rm s}$) &
      $\omega_{{\rm LO},m}$ (${\rm rad}/{\rm s}$) & $\gamma_m$ (${\rm rad}/{\rm s}$) \\
      \hline
      $m=\perp$ & 4.87 & $2.58\times 10^{14}$ & $3.03\times 10^{14}$ & $9.42\times
      10^{11}$ \\
      \hline
      $m=\parallel$ & 2.95 & $1.47\times 10^{14}$ & $1.56\times 10^{14}$ & $7.54\times
      10^{11}$ \\
      \hline
    \end{tabular}
    \caption{\label{tab:1}Parameters for calculating the dielectric tensor of hBN. }
  \end{center}
\end{table}

\begin{figure}
\centering
\includegraphics[width=\columnwidth]{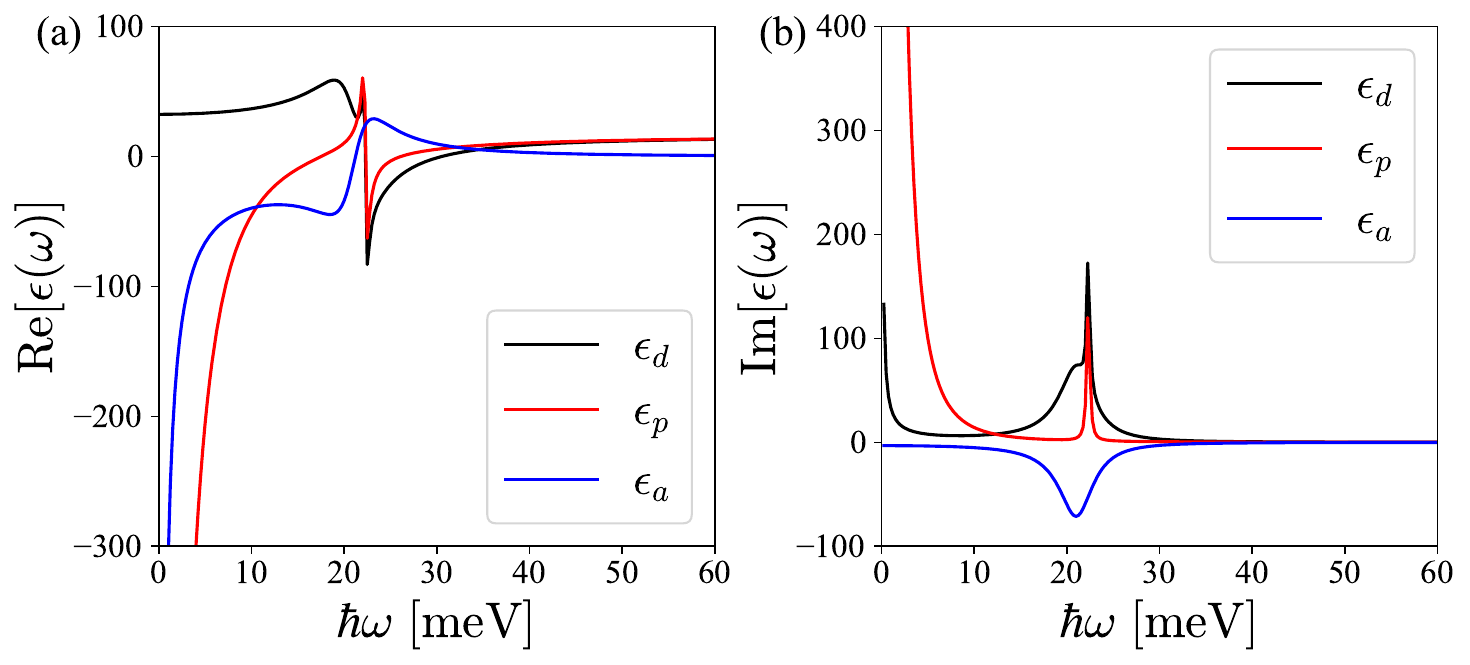} \\
\caption{(a) Real and imaginary parts of the components of the dielectric tensor of InSb
$\bar{\bar{\epsilon}}_{\rm MO}(\omega)$ at $B=1\,$T. }
\label{figS2}
\end{figure}

\section*{Magneto-optic medium}
We consider magneto-optic medium InSb where the magnetic field $B$ is applied along the
$y$ direction. From the Drude model, the dielectric tensor is expressed as~\cite{InSb_76}
\begin{equation}
  \bar{\bar{\epsilon}}_{\rm MO}(\omega) = 
  \begin{bmatrix}
    \epsilon_d & 0 & i\epsilon_a \\
    0 & \epsilon_p & 0 \\
    -i\epsilon_a & 0 & \epsilon_d
  \end{bmatrix} ,
\end{equation}
where 
\begin{align}
  \epsilon_d &= \epsilon_{\infty} \bigg[ 1+ \frac{\omega_L^2 -\omega_T^2}{\omega_T^2
  -\omega^2 -i\Gamma\omega} + \frac{\omega_p^2(\omega+i\gamma)}{\omega\big[
  \omega_c^2-(\omega+i\gamma)^2 \big]} \bigg], \\
  \epsilon_p &= \epsilon_{\infty} \bigg[ 1+ \frac{\omega_L^2 -\omega_T^2}{\omega_T^2
  -\omega^2 -i\Gamma\omega} - \frac{\omega_p^2}{\omega(\omega+i\gamma)} \bigg], \\
  \epsilon_a &= \frac{\epsilon_{\infty}\omega_p^2\omega_c}{\omega\big[ (\omega+i\gamma)^2
  -\omega_c^2 \big]} .
\end{align}
The parameters are taken from Ref.~\cite{InSb_76} with
the high-frequency dielectric constant $\epsilon_\infty = 15.7$, 
the longitudinal optical phonon frequency $\omega_L = 3.62\times 10^{13}\,{\rm rad/s}$, 
the transverse optical phonon frequency $\omega_T = 3.39\times 10^{13}\,{\rm rad/s}$, 
the phonon damping constant $\Gamma = 5.65\times 10^{11}\,{\rm rad/s}$, 
the free-carrier damping constant $\gamma = 3.39\times 10^{12}\,{\rm rad/s}$, 
the plasma frequency $\omega_p = 3.14\times 10^{13}\,{\rm rad/s}$, and
the cyclotron frequency $\omega_c = 8.02\times 10^{12}\,{\rm rad/s}$ for $B=1\,$T.
The real and imaginary parts of the components of $\bar{\bar{\epsilon}}_{\rm MO}(\omega)$
are shown in Figs.~\ref{figS2}(a) and \ref{figS2}(b), respectively.

The dispersion of the surface polaritons in the Voigt configuration can be
obtained as~\cite{Chiu1972, BinHu15, GT_WSM}
\begin{equation} \label{SPP}
  \epsilon_v \beta_0 +\beta_2 -i\epsilon_a q_x /\epsilon_d =0 ,
\end{equation}
with the Voigt dielectric function $\epsilon_v = \epsilon_d-\epsilon_a^2/\epsilon_d$. 
The out-of-plane wave vectors in air and in the magneto-optic medium are, respectively,
given by $\beta_0 = \sqrt{k_0^2 - q_x^2}$ and $\beta_2 = \sqrt{\epsilon_v k_0^2 - q_x^2}$.
One can solve Eq.~\eqref{SPP} to get
\begin{align}
  & \big[(\epsilon_a^2/\epsilon_d^2-\epsilon_v^2-1)^2 -4\epsilon_v^2\big] q_x^4 \notag \\
  +& 2(\epsilon_v^2+\epsilon_v)(\epsilon_a^2/\epsilon_d^2-\epsilon_v^2-1+2\epsilon_v) k_0^2
  q_x^2 \notag \\
  +& (\epsilon_v^2-\epsilon_v)^2 k_0^4 = 0 .
\end{align}
In the absence of a magnetic field, the dispersion becomes
$q_x=\pm \sqrt{\epsilon_d/(\epsilon_d+1)}k_0$. 
The calculation details of the reflection coefficient $\rho_2$ can be found in the
Supporting Information of Ref.~\cite{GT_WSM}.

\end{document}